# Adsorption of Oxygen Molecules on Individual Carbon Single-walled Nanotubes

A. Tchernatinsky[†], B. Nagabhirava[‡], S. Desai[†], G. Sumanasekera[†], B. Alphenaar[‡], C.S. Jayanthi[†], and S.Y. Wu[†*]

[†]Department of Physics, University of Louisville

[‡]Department of Electrical and Computer Engineering, University of Louisville

Louisville, KY 40292

*Corresponding author. E-mail: sywu0001@gwise.louisville.edu

ABSTRACT

Our study of the adsorption of oxygen molecules on individual semiconductiong single-walled carbon nanotubes at ambient conditions reveals that the adsorption is physisorption, that the resistance without $O_2$ increases by ~two orders of magnitude as compared to that with $O_2$, and that the sensitive response is due to the pinning of the Fermi level near the top of the valence band of the tube resulting from impurity states of $O_2$ appearing above the valence band.

Interest in gas adsorption by carbon nanotubes at ambient conditions has been spurred by the demonstrations of the potential of single-walled carbon nanotube (SWCNT)-based gas sensors[1-3], specifically, the establishment of the interdependence between gas adsorption and transport properties of CNTs. In recent years, experimental studies on the adsorption of oxygen molecules by SWNT-bundles or mats included the measurements of electrical resistance and thermoelectric power[2,3], the effect of adsorption of $O_2$ on the barrier of metal-semiconductor contact[4,5], and the kinetics of $O_2$ adsorption and desorption[6]. The picture emerged from these studies relevant to gas sensing indicates that the electrical resistance changes by about 15% between gassing and degassing[2], that the hole doping of semiconducting SWNT (s-SWNT) in air is by the adsorption of $O_2$ in the bulk of s-SWNTs[5] rather than at the contact[4], and that the adsorption of $O_2$ has the characteristics of physisorption[6]. Theoretical investigations of the adsorption of $O_2$ on SWNTs have also been carried out, using spin-unpolarized as well as spin-polarized density functional theory (DFT) methods[7-12]. Studies of the adsorption of $O_2$ on small-diameter (8,0) SWNT based on the spin-unpolarized DFT within the local density approximation (LDA) predicted a weak hybridization between states of $O_2$ and those of the s-SWNT with an estimated charge transfer of *~0.1e*[7,9], leading to a binding of $O_2$ at a distance less than 3 Å from the s-SWNT. The hole doping of the s-SWNT was attributed to the pinning of the Fermi level at the top of the valence band due to the adsorption of $O_2$. With the $O_2$ molecule having a triplet ground state, the more realistic calculations based on the spin-polarized gradient-corrected DFT[8,10,12], on the other hand, yielded a very weak bonding at ~4 Å with no significant charge transfer, indicating that an $O_2$ molecule in the more stable triplet state is only physisorbed on a s-SWNT. For the triplet state of $O_2$ adsorbed on the (8,0) SWNT, two degenerated *pp$\pi$\** bands were found to split into four bands, with the two un-occupied $pp\pi^*(\downarrow)$ bands rising ~0.35 eV above the top of the valence band at the $\Gamma$ point[12], casting some doubt about the hole doping picture deduced from the un-polarized calculation.

In order to obtain a coherent and consistent picture of the adsorption of $O_2$ by individual s-SWNTs, we have conducted a careful experimental and theoretical investigation of the adsorption of $O_2$ molecule by individual SWNTs to shed light on the nature of adsorption and its effect on the transport properties of SWNTs. Experimentally, contacts were made to a few very dispersed SWNTs using e-beam lithography. The experiment was first conducted under ambient conditions in air (room temperature and atmosphere pressure). The resistance was monitored during each exposure to air and subsequent pumping ($10^{-6}$ torr). A resistance change of more than one order of magnitude was observed for the first time as a result of the adsorption of $O_2$ by *individual* s-SWNTs, in dramatic contrast to a mere 15% change observed for SWNT bundles or mats. Furthermore, the onset of the change in resistance was in minutes. These observations clearly demonstrated the feasibility of constructing s-SWNT-based chemical sensors. To be more consistent with the experimental result, we have carried out a study on the adsorption of an $O_2$ molecule by a larger SWNT than the one considered in previous studies, the (14,0) SWNT, that is closer to the range of diameters in the experiment, using the spin-polarized DFT method. Our calculation (spin-polarized GGA) yielded a shallow potential well of depth of the order of ~0.05 eV at ~3.6 Å from the surface of the SWNT, consistent with the picture of physisorption. We have determined the pinning of the Fermi energy due to the impurity level associated with $O_2$. Our estimate of the resistance based on the result of the (14,0) tube with the adsorption of the $O_2$ molecule is in excellent agreement with the observed initial resistance in air, indicating the metallization of the s-SWNT by hole doping associated with the physisorbed $O_2$. We have also predicted a change in the resistance about two orders of magnitude between gassing and degassing.

Individual SWNTs were synthesized using CVD with Fe catalyst and $CH_4$ on a $SiO_2$/Si substrate with pre patterned grid marks. Silicon (100) with a thin oxide layer (0.4 μm) was selected for the growth process. Grid pattern (Au alignment marks) was fabricated on the $SiO_2$/Si substrate using e-beam lithography and etched using BOE. The alignment marks were etched so that they can be seen in atomic force microscopy (AFM) imaging. The preparation of catalyst solution follows the procedure given in Ref. 13. Fe nanoparticles were dispersed on the substrate from $Fe(NO_3)_3$ propanol solution. After washing with hexane, the substrate was loaded into the CVD reactor and heated to 900 $^0$C in flowing Ar/$H_2$ (100 sccm of 10% $H_2$ in Ar). After reduction at 900$^0$C for 10 minutes, Methane, the carbon feed, was introduced at a rate of 400 sccm for 2 minutes. The sample was cooled in Argon. The sample SWNTs were imaged using the AFM with reference to the alignment marks in the grid pattern and the Au/Ti contacts were made to the SWNTs using e-beam lithography and evaporation. Larger contact pads were deposited on the e-beam defined contacts using optical lithography (see Fig. 1). The device was loaded into a quartz reactor equipped with a turbo-molecular pump capable of evacuating to $10^{-7}$ Torr for in situ studies. The reactor has provisions for gases and chemical vapors.

The experiment was first conducted under ambient conditions (room temperature and atmosphere pressure). Two-probe resistance of the device was measured during the exposure to the air and subsequent pumping at room temperature. The resistance was continuously monitored during each exposure and subsequent pumping. Figure 2 shows the time evolution of the 2-probe resistance of the device during pumping and subsequent exposure. Data for two cycles are shown. The 2-terminal resistance of the as prepared device was ~300 k$\Omega$. During pumping ($\leq 10^{-6}$ Torr), the resistance started to increase and eventually saturated at a value of ~16 M$\Omega$ within a period of ~1 hr. This represents a change of the resistance of close to two orders of magnitudes for *individual* SWNTs, a dramatic change in comparison to the ~15% change observed for SWNT bundles or mats[2]. When exposed to air at this point, the resistance started to decrease, initially with an abrupt drop to ~ 2.5 M$\Omega$ within ~ 15 min. This substantial drop in the resistance within such a short time interval after the exposure to the air indicates the sensitivity of the response of *individual* SWNTs to the absorption of gases in air. The initial drop in resistance was followed by a much slower decrease, saturating at the initial value of ~300 k$\Omega$ in ~ 10 hrs. Similar behavior was observed in another cycle as shown in Fig. 2. The experimental findings suggest that the fabricated device is most likely composed of s-SWNTs and that the findings reflect the response of the transport properties of s-SWNT during the exposure to $O_2$ in air and subsequent pumping. We established the semiconducting nature of our device by measuring the gate voltage dependence of the conductance of the device at room temperature, using Si substrate as the back gate. We found that when the positive gate voltage is increased, the conductance decreases while conductance increases when the negative gate voltage is increased. As the conductance of metallic tubes should have little or no gate voltage dependence, and on the other hand, increasing negative gate voltage adds more holes to s-SWNTs, thereby increasing the conductance, we conclude that our device consists of only s-SWNTs.

To shed light on the physics underlying the change in the transport properties during the adsorption and desorption of $O_2$ molecules by s-SWNTs, we carried out a detailed study of the adsorption of an $O_2$ molecule on a (14,0) s-SWNT, using spin-polarized local density approximation (LDA) as well as spin-polarized generalized gradient approximation (GGA) DFT-methods in the Vienna *ab initio* simulation package (VASP)[14-16]. We chose to use the (14,0) SWNT as the benchmark because its diameter (*d*=1.10 nm) is close to the range of diameters of typical SWNTs and a recent DFT calculation has established the *1/d*-dependence of the energy gap of s-SWNTs to be valid only for $d \geq 1.0$ nm[17]. In our calculation, we used a supercell of the size 26x26x8.54 (Å) to cut down the potential image effect between SWNTs. Along the axial direction of the SWNT, this supercell consists of two SWNT unit cells so that the calculation reflects well the situation of the physisorption of individual $O_2$ molecules. Vanderbilt ultrasoft pseudopotential[18,19] and Perdew and Zunger functional[20], with the GGA correction of Perdew *et al*[21], were used for the self-consistent spin-polarized solution. The energy cut-off was set at 700 eV. Monkhorst-Pack scheme with 1x1x11 *k*-point mesh was used for sampling the Brillouin zone. Full optimization of the structural configuration of SWNT+$O_2$ and the lattice constants were carried out using the conjugate gradient method with the energy convergence of *$10^{-5}$* eV and forces $\leq 10^{-2}$ eV/Å.

Our calculations, using spin-unrestricted LDA as well as GGA, confirmed that the triplet $O_2$ state has the lower energy as compared to the singlet state. Before using the VASP code to investigate the benchmark case of the (14,0) s-SWNT+$O_2$, we applied it to the case of the (8,0) s-SWNT+$O_2$ with $O_2$ near the T site[12]. The optimization yields a result in excellent agreement with the corresponding result in Ref. 12 (see Fig. 5g in Ref. 12). Having established the validity of the VASP code, we carried out optimizations of the adsorption of $O_2$ on the (14,0) s-SWNT with spin-polarized methods (LDA & GGA). We found the binding to be the strongest for the triplet $O_2$ molecule near the top of two adjacent zigzag bonds (T site), with the molecular axis perpendicular to the axial direction of the SWNT (see Fig. 3). The relaxed bonding geometries (bond length and equilibrium orientation of $O_2$) from both methods are almost the same except the equilibrium distance from $O_2$ to the surfaces of the (14,0) SWNT. Figure 3 shows a weak potential well of depth ~0.1 eV at a distance of ~3.0 Å for the LDA result and a very shallow well of ~0.03 eV at a distance of ~3.5 Å for the GGA result. These results are consistent with the scenario of physisorption. For physisorption characterized by weak interactions, LDA tends to overestimate the binding and underestimate the equilibrium distance while GGA tends to underestimate the binding. From our result, one can conclude that the phsisorption of $O_2$ on the (14,0) s-SWNT is characterized by a potential well of depth between 0.03 and 0.1 eV and an equilibrium distance between 3.0 and 3.5 Å.

Figure 4 shows the band structures in the vicinity of the energy gap of relaxed configurations of the adsorption of triplet $O_2$ on the surface of the (14,0) s-SWNT obtained by LDA and GGA respectively. The energy gap obtained by the GGA calculation is ~0.69 eV while that by LDA is ~0.60 eV. We have also checked the band structure for the pristine (14,0) s-SWNT, using same methods. We found the same values for the gap and no difference in the band structures as compared to those for the case of (14,0)+$O_2$ by the respective method. Furthermore, the unoccupied oxygen $pp\pi^*(\downarrow)$ bands were found to appear within the gap of the s-SWNT almost dispersionless in both calculations. These results present an unambiguous indication of a very weak interaction between the oxygen molecule and the (14,0) s-SWNT, reinforcing the scenario of the physisorption. In this sense, our calculations have essentially established the placement of the empty impurity bands within the gap of the s-SWNT. Specifically, for the GGA calculation, the lower $pp\pi^*(\downarrow)$ band is ~0.20 eV above the top of the valence band while that for the LDA is ~0.24 eV above the top of the valence band. To summarize, our study indicates no charge transfer between $O_2$ and the (14,0) s-SWNT. The effect of the presence of the oxygen impurity bands is to pin the Fermi level to the vicinity of the top of the valence band.

The conductance for the pristine s-SWNT and that for the s-SWNT with $O_2$ under ambient conditions can be estimated according to

$$G = \frac{2e^2}{h} \int_{-\infty}^{\infty} T(E)(-\frac{\partial f}{\partial E})dE \approx G_0[\frac{2}{1+e^{\Delta/2kT}}]$$

(1)

where $T(E)$ is the transmission coefficient as a function of $E$ and may be approximated by $2$ in the vicinity of the Fermi energy for SWNTs, $f(E)$ the Fermi distribution function, $G_0 = 4e^2/h$, the quantum conductance, and $\Delta$ the energy gap. Using Eq. (1) based on the energy gap with or without $O_2$ obtained by LDA as well as GGA, we have calculated the resistances of the (14,0) s-SWNT with or without $O_2$ at room temperature. The results are shown in Table 1. It can be seen that the GGA method yields a value of ~400 k$\Omega$ for the resistance with $O_2$, in very good agreement with the experimental result, while the LDA method gives rise to a value of ~680 k$\Omega$. For the (14,0) s-SWNT, the GGA method leads to a resistance increase by a factor of $\sim 1.33 \times 10^4$ between the resistance of the s-SWNT without $O_2$ and that with $O_2$ while the LDA an increase by a factor of $1.08 \times 10^3$. This resistance change can be attributed to the pinning of the Fermi level to the vicinity of the valence band due to the presence of the empty oxygen bands. Since the diameter of a typical SWNT is ~1.40 nm and the gap follows a *1/d*-dependence on the diameter for $d \geq 1$ nm, we estimated the energy gap of the typical s-SWNT using the calculated gap of the (14,0) SWNT ($d$=1.09 nm) according to $\Delta_{adj} = \Delta \times 1.09/1.4$. Using $\Delta_{adj}$, we obtained a resistance increase by a factor of 82 for the LDA result and 652 for the GGA result, consistent with the experimental result.

Our experimental study and theoretical analysis clearly lead to the following conclusion concerning the adsorption and desorption of $O_2$ molecules on s-SWNTs. (i) The resistance change between the desorption and the absorption of $O_2$ molecules by an *individual* s-SWNT is ~two orders of magnitude. The response of *individual* s-SWNTs to the exposure to $O_2$ molecules is therefore far more sensitive as compared to the response of SWNT bundles or mats studied previously. (ii) The adsorption of $O_2$ molecules on s-SWNTs is unequivocally physisorption. There is no charge transfer between the $O_2$ molecules and the s-SWNT. (iii) The sensitive response of s-SWNTs to the adsorption of $O_2$ molecules is due to the pinning of the Fermi level near the top of the valence band.

We would like to acknowledge the support by the NSF (DMR-0112824 and ECS-0224114) and the DOE (DE-FG02-00ER4582).

Table 1. Resistances of the (14,0) SWNT without and with $O_2$ at room temperature, and their ratio, calculated by spin-polarized LDA and GGA respectively. Also shown are the adjusted resistances corresponding to a s-SWNT (without $O_2$) with a diameter of 1.40 nm.

|     | R w/o $O_2$(**kΩ**) | R with $O_2$(**kΩ**) | Ratio | R w/o $O_2$(**kΩ**) adjusted | Ratio Adjusted |
| --- | --- | --- | --- | --- | --- |
| LDA | $7.4 \cdot 10^5$ | $6.8 \cdot 10^2$ | $1.08 \cdot 10^3$ | $5.6 \cdot 10^4$ | $8.20 \cdot 10^1$ |
| GGA | $5.2 \cdot 10^6$ | $4.0 \cdot 10^2$ | $1.33 \cdot 10^4$ | $2.6 \cdot 10^5$ | $6.52 \cdot 10^2$ |

FIGURES

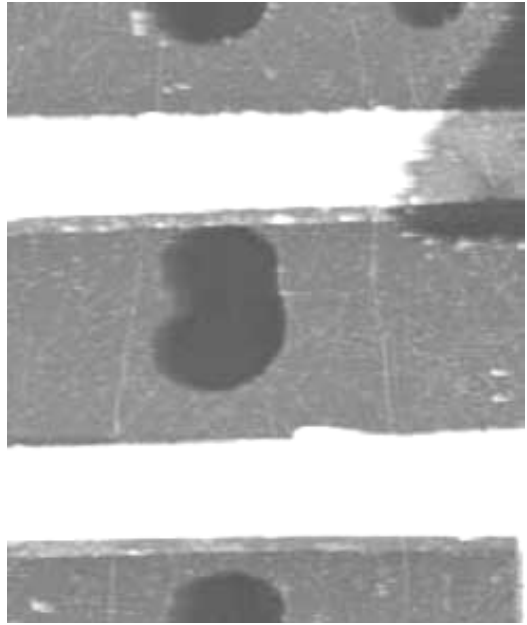

Figure 1. An AFM image of two SWNTs buried underneath Au/Ti contacts.

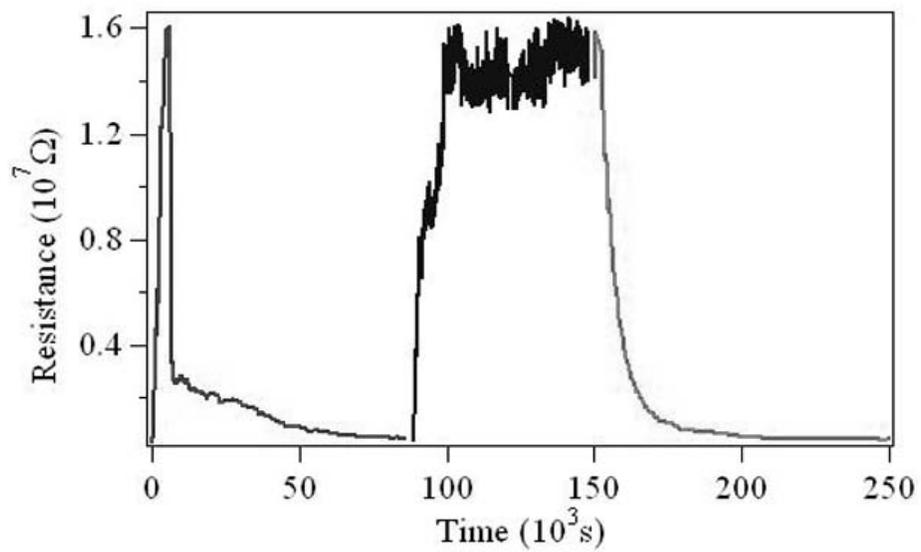

Figure 2. Two cycles of the time evolution of the 2-probe resistance of the sample during pumping and subsequent exposure to air.

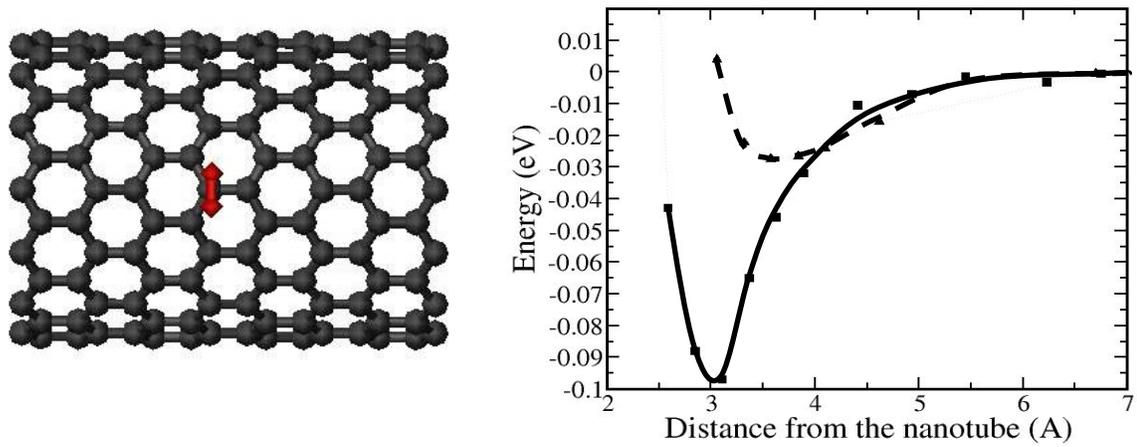

Figure 3. The binding of a triplet O$_2$ near the top of two adjacent zigzag bonds. Left panel: The equilibrium configuration. Right panel: Energy *vs*. distance curves (dotted curve, GGA result; solid curve, LDA result).

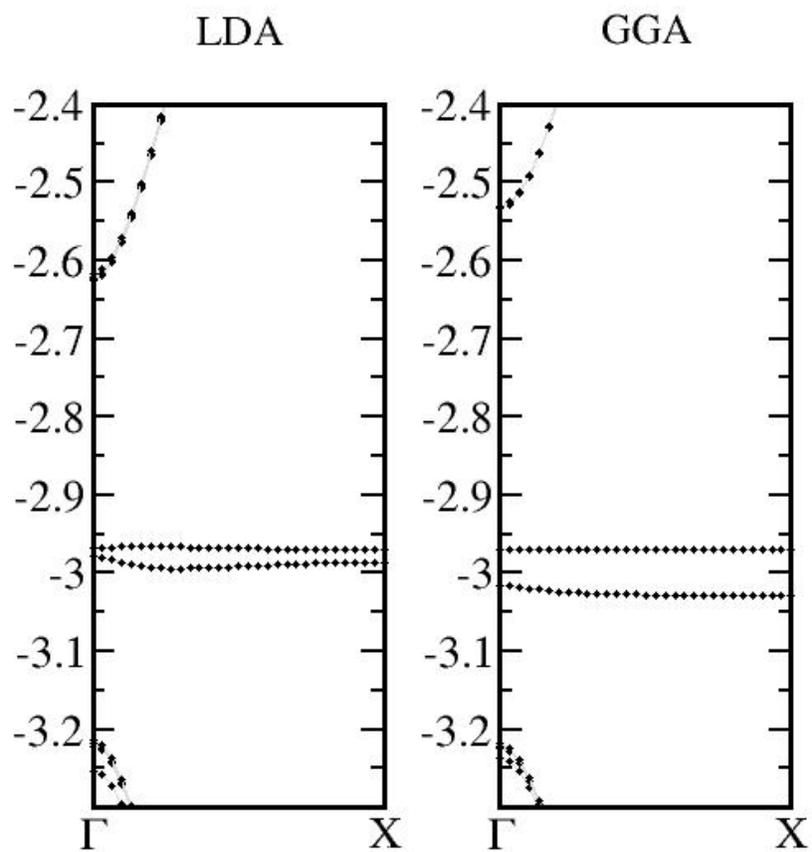

Figure 4. Band structures corresponding to the equilibrium configurations. Left panel: LDA result. Right panel: GGA result.

References

[1] J. Kong, N.R. Franklin, C. Zhou, M.G. Chapline, S. Peng, K. Cho, and H. Dai, *Science* **287**, 622 (2000).

[2] P.G. Collins, K. Bradley, M. Ishigami, and A. Zettl, *Science* **287**, 1801 (2000).

[3] C.K.W. Adu, G.U. Sumanasekera, B.K. Pradham, H.E. Romero, and P. C. Eklund, *Chem. Phys. Lett.* **337**, 31 (2001).

[4] V. Derycke, R. Martel, J. Appenzeller, and Ph. Avouris, *Appl. Phys. Lett.* **80**, 2773 (2002).

[5] M. Shim, and G.P. Siddons, *Appl. Phys. Lett.* **83**, 3564 (2003).

[6] H. Ulbricht, G. Moos, and T. Hertel, *Phys. Rev. B* **66**, 075404 (2002).

[7] S.-H. Jhi, S.G. Louie, and M.L. Cohen, *Phys. Rev. Lett.* **85**, 1710 (2000).

[8] D.C. Sorescu, K.D. Jordan, and Ph. Avouris, *J. Phys. Chem B* **105**, 11227 (2001).

[9] J. Zhao, A. Buldum, J. Han, and J.P. Lu, *Nanotechnology* **13**, 195 (2002).

[10] P. Giannozzi, R. Car, and G.J. Scoles, *Chem. Phys.* **118**, 1003 (2003).

[11] M. Grujicic, G. Cao, and R. Singh, *Appl. Surface Science* **211**, 166 (2003).

[12] S. Dag, O. Gülseren, T. Yildirim, and S. Ciraci, *Phys. Rev. B* **67**, 165424 (2003).

[13] J.H. Hafner, C.L. Cheung, Th. Oosterkamp, and C.M. Lieber, *J. Phys. Chem. B* **105**, 743 (2001).

[14] G. Kresse, and J. Hafner, *Phys. Rev. B* **48**, 13115 (1993).

[15] G. Kresse, and J. Furthmuller, *Phys. Rev. B* **54**, 11169 (1996).

[16] G. Kresse, and J. Furthmuller, *Comput. Mat Sci.* **6**, 15 (1996).

[17] V. Zólyomi, and J. Kürti, *Phys. Rev. B* **70**, 085403 (2004).

[18] D. Vanderbilt, *Phys. Rev. B* **41**, 7892 (1990).

[19] G. Kresse, and J. Hafner, *J. Phys.: Cond. Matter* **6**, 8245 (1994).

[20] J.P. Perdew, and A. Zunger, *Phys. Rev. B.* **23**, 5048 (1981).

[21] J.P. Perdew, J.A. Chevary, S.H. Vosko, K.A. Jackson, R.D. Rederson, D.J. Singh, and C. Fiolhais, *Phys. Rev. B* **46**, 6671 (1992).